\documentclass[10pt, a4paper]{article}
\usepackage{lrec}
\usepackage{graphicx}
\usepackage{amsmath,graphicx}
\usepackage[dvipsnames,svgnames,x11names,table,xcdraw]{xcolor}
\usepackage{multirow}
\usepackage{url}
\usepackage{booktabs}
\usepackage[normalem,normalem,normalbf]{ulem}
\usepackage{tabularx}
\usepackage{soul}
\usepackage{tikz}
\usepackage{epstopdf}
\usepackage[latin1]{inputenc}

\usepackage{hyperref}
\usepackage{xstring}

\newcommand{\secref}[1]{\StrGobbleRight{\getrefnumber{#1}}{1}}

\title{\textbf{Semi-supervised Development of ASR Systems for Multilingual Code-switched Speech in Under-resourced Languages}}

\name{Astik Biswas$^1$, Emre Y\i lmaz$^2$, Febe de Wet$^1$, Ewald van der Westhuizen$^1$, Thomas Niesler$^{1}$}

\address{$^1$  Department of Electrical and Electronic Engineering, Stellenbosch University, South Africa \\
         $^2$ Department of Electrical and Computer Engineering, National University of Singapore, Singapore \\
         \{abiswas, fdw, ewaldvdw, trn\}@sun.ac.za, emre@nus.edu.sg}

\abstract{
This paper reports on the semi-supervised development of acoustic and language models for under-resourced, code-switched speech in five South African languages.
Two approaches are considered.
The first constructs four separate bilingual automatic speech recognisers (ASRs) corresponding to four different language pairs between which speakers switch frequently.
The second uses a single, unified, five-lingual ASR system that represents all the languages (English, isiZulu, isiXhosa, Setswana and Sesotho).
We evaluate the effectiveness of these two approaches when used to add additional data to our extremely sparse training sets.
Results indicate that batch-wise semi-supervised training yields better results than a non-batch-wise approach.
Furthermore, while the separate bilingual systems achieved better recognition performance than the unified system, 
they benefited more from pseudo-labels generated by the five-lingual system than from those generated by the bilingual systems.\\[-2mm] \newline  } 

\begin{document}

\maketitleabstract

\section{Introduction}
\label{sec:intro}
Much research has already been dedicated to the development of automatic speech recognition (ASR) for code-switching between various languages.
English-Mandarin has probably been studied most extensively~\cite{li2013improved,li2013language,zeng2018end,vu2012first,taneja2019exploiting}, but other language pairs such as Frisian-Dutch \cite{yilmaz2018semi}, Hindi-English \cite{pandey2018phonetically,emond2018transliteration}, English-Malay \cite{ahmed2012automatic} and French-Arabic \cite{amazouz2019addressing} have also received some attention.
Developing models that are robust to the additional complexity associated with code-switching is challenging.
The task becomes even more difficult when the languages in question are under-resourced since small text and acoustic data sets limit modelling capacity.

South Africa has a multilingual population of 57 million people and 11 official languages, including English.
Due to the variety of geographically co-located languages, code-switching - the alternation between languages during communication - is a common phenomenon.
Code-switching is most prevalent between English, a highly-resourced language,  and the South African Bantu languages,  which  are  all under-resourced.
A corpus of code-switched speech originating from South African soap operas has recently been compiled to enable the development of ASR for this type of speech \cite{van2018city}.

Previous work demonstrated that multilingual training using in-domain soap opera code-switched speech \cite{biswas2018IS,yilmaz2018building} and poorly matched monolingual South African speech \cite{biswas2018improving} improves the performance of both bilingual and five-lingual ASR systems when the additional training data is from a closely-related language. Specifically, isiZulu, isiXhosa, Sesotho and Setswana belong to the same Bantu language family and were found to complement each other when combined into a multilingual training set.  
It has also been shown that, in comparison with in-domain code-switched data, out-of-domain monolingual speech yields relatively little performance improvement in acoustic modelling~\cite{biswas2018improving}. 
However,  the in-domain training data that is currently available remains insufficient for robust ASR development and hence obtaining more in-domain data remains key to improve the recognition accuracy of code-switched speech.

Compiling a multilingual corpus of code-switched speech is an extremely labour intensive process, both in terms of effort and time, because manual segmentation and annotation of the data are required. 
In the absence of manually annotated material, automatically transcribed training material has been shown to be useful in under-resourced scenarios using semi-supervised training ~\cite{thomas2013deep,yilmaz2018semi,Guo2018}.
For example, this strategy was used successfully to obtain bilingual and five-lingual ASR systems using 11.5 hours of manually segmented but untranscribed soap-opera speech~\cite{biswas2019IS2}.
Furthermore, the bilingual systems trained with automatic transcriptions generated by the five-lingual transcription system achieved the best performance. 

Motivated by these results, we now investigate a batch-wise semi-supervised technique in which we incorporate additional batches of manually segmented but untranscribed soap opera data for acoustic and language modelling.
Initial transcriptions were generated using our best systems trained on the manually transcribed speech.
Given the multilingual nature of the data, the transcription systems must not only provide the orthography, but also the language(s) present at each location in the segment.
Each utterance was therefore presented to the four individual code-switching systems as well as to the five-lingual system.
In both cases two training configurations were considered, the first presenting all the data in one pass and the second presenting the data in smaller batches.

Finally, we also present language modelling experiments that investigate the inclusion of the automatically generated transcriptions and artificially generated text as training material for English-isiZulu.

\section{Multilingual soap opera corpus}
\label{SEC:corpus}

This work uses a multilingual corpus including examples of code-switching between South African English and four Bantu languages. 
The corpus, which was compiled from South African soap opera episodes, contains 21 hours of annotated South African code-switched speech data divided into four language pairs: English-isiZulu (EZ), English-isiXhosa (EX), English-Setswana (ET), and English-Sesotho (ES).
Of the Bantu languages, isiZulu and isiXhosa belong to the Nguni language family while Setswana and Sesotho are Sotho-Tswana languages. The speech in question is typically fast and often expresses emotion. 
These aspects of the data in combination with the high prevalence of code-switching makes it a challenging corpus for ASR experiments. 

The corpus is however still under construction.
During the first phase of development, more than 600 South African soap opera episodes were manually segmented and a substantial portion of this also manually transcribed.
The second phase is currently underway and has thus far contributed manually segmented but still untranscribed data to the corpus.

\subsection{Manually transcribed data (ManT)}
\label{SEC:corpus:transcribed}
The version of the soap opera corpus we used to develop our first code-switching ASR systems consisted of 14.3 hours of speech divided into four language-balanced sets, as described in~\cite{van2018city}. 
In addition to the language-balanced sets, approximately 9 hours of manually transcribed monolingual English soap opera speech was also available.
This data was initially excluded to avoid a bias toward English.
However, pilot experiments indicated that, counter to expectations, its inclusion enhanced recognition performance.
These 9 hours of English data were therefore merged with the balanced sets for the experiments described here. 

\begin{table}[t]
\centering 
\renewcommand*{\arraystretch}{1.0}
\resizebox{\columnwidth}{!}{
\begin{tabular}{c r r r r r r}
\toprule
{\bf{Language}} & \begin{tabular}[c]{@{}c@{}}\bf{Mono}\\  (m)\end{tabular} & \begin{tabular}[c]{@{}c@{}}\bf{CS}\\ (m)\end{tabular}&
\begin{tabular}[c]{@{}c@{}}\bf{Total}\\ (h)\end{tabular}&
\begin{tabular}[c]{@{}c@{}}\bf{Total}\\ (\%)\end{tabular}&
\begin{tabular}[c]{@{}c@{}}\textbf{Word}\\ \textbf{tokens}\end{tabular} & \begin{tabular}[c]{@{}c@{}}\textbf{Word} \\ \textbf{types}\end{tabular} \\ \midrule
English & 755.0              & 121.8  & 14.6 & 69.3 & 194\,426              & 7\,908              \\
isiZulu & 92.8               & 57.4  & 2.5 & 11.9 & 24\,412              & 6\,789              \\
isiXhosa & 65.1               & 23.8  & 1.5 & 7.0  & 13\,825              & 5\,630              \\ 
Sesotho & 44.7               & 34.0  & 1.3 & 6.2  & 22\,226              & 2\,321              \\ 
Setswana & 36.9               & 34.5  & 1.2 & 5.6  & 21\,409              & 1\,525              \\ \midrule
{\bf{Total}} & 994.5              & 271.5  & 21.1 & 100.0  & 276\,290              & 24\,170              \\ \bottomrule
\end{tabular}%
}
\caption{Duration in minutes (min) and hours (h) as well as word type and token counts for the unbalanced training speech.}
\label{tab:duration_unbalanaced_corpora}
\end{table}

The composition of the unbalanced training speech is reported in Table~\ref{tab:duration_unbalanaced_corpora}. 
An overview of the statistics for the development (Dev) and test (Test) sets for each language pair is given in Table \ref{tab:corpora_stat}.
The table includes values for the total duration as well as the duration of the monolingual  and code-switched  segments. 
The test sets contain no monolingual data.
A total of approximately 4,000 language switches (English-to-Bantu and Bantu-to-English) are observed in the test set.

The number of unique English, isiZulu, isiXhosa, Sesotho and Setswana words in the corpus are 8\,275, 11\,352, 6\,169, 2\,792, 1\,902 respectively.
IsiZulu and isiXhosa have relatively large vocabularies due to their agglutinative nature. 
This property adds to the challenge of developing accurate ASR systems in these languages.

\begin{table}[t]
\small
	\centering
    \renewcommand*{\arraystretch}{0.9}
		\begin{tabular*}{0.47\textwidth}{@{\extracolsep{\fill}}c r r r r r @{}}
			\toprule
			\multicolumn{6}{c}{\textbf{English-isiZulu}} \\
			& emdur & zmdur & ecdur & zcdur & \textbf{Total} \\
			\textbf{Dev} & 0.00 & 0.00 & 4.01 & 3.96 & 8.00 \\
			\textbf{Test} & 0.00 & 0.00 & 12.76 & 17.85 & 30.40 \\ \midrule
			\multicolumn{6}{c}{\textbf{English-isiXhosa}} \\
			& emdur & xmdur & ecdur & xcdur & \textbf{Total} \\
			\textbf{Dev} & 2.86 & 6.48 & 2.21 & 2.13 & 13.68 \\
			\textbf{Test} & 0.00 & 0.00 & 5.56 & 8.78 & 14.34 \\ \midrule
			\multicolumn{6}{c}{\textbf{English-Setswana}} \\
			 & emdur & tmdur & ecdur & tcdur & \textbf{Total} \\
			\textbf{Dev} & 0.76 & 4.26 & 4.54 & 4.27 & 13.83 \\
			\textbf{Test} & 0.00 & 0.00 & 8.87 & 8.96 & 17.83 \\ \midrule
			\multicolumn{6}{c}{\textbf{English-Sesotho}} \\
			 & emdur & smdur & ecdur & scdur & \textbf{Total} \\
			\textbf{Dev} & 1.09 & 5.05 & 3.03 & 3.59 & 12.77 \\
			\textbf{Test} & 0.00 & 0.00 & 7.80 & 7.74 & 15.54 \\ \bottomrule
		\end{tabular*}%
	\caption{Duration (minutes) of English, isiZulu, isiXhosa, Sesotho, Setswana monolingual (mdur) and code-switched (cdur) utterances in the code-switching development and test sets.}
	\label{tab:corpora_stat}

\end{table}

\subsection{Manually segmented data: Batch 1}
\label{SEC:corpus:untranscribed1}
A set of approximately 11 hours of manually segmented speech representing 127 different speakers was produced in addition to the manually transcribed data introduced in the previous section.
Segmentation was performed manually by experienced language practitioners.
This data set (\textbf{B1}) was automatically transcribed during our initial investigations into semi-supervised acoustic model training~\cite{biswas2019IS2}.
Two sets of automatic transcriptions derived from B1 are considered: one obtained using four bilingual systems (AutoT\_B$_{\rm B1}$) and the other using a five-lingual system (AutoT\_F$_{\rm B1}$).

\subsection{Manually segmented data: Batches 2 \& 3}
\label{SEC:corpus:untranscribed2}
A subsequent phase of corpus development, currently still underway, has produced two new batches of manually segmented data.
Manual transcriptions of this data are not yet available.
These data sets will be referred to as Batch 2 (\textbf{B2}) and Batch 3 (\textbf{B3}), respectively.
In contrast to B1, the segmentation was done by trained assistants because no specialist language practitioners were available.
Hence, the quality of the segments in B2 and B3 may differ from those in B1. 

Batch B2 includes approximately 24 hours of speech produced by 157 speakers, while B3 contains a further 30 hours of speech from 145 speakers.
Most speakers occur in both batches and the languages spoken in the segments are not labelled.
South African languages other than the five present in the transcribed data are known to occur in these batches, but to a limited extent.
%
%
\section{Acoustic Modelling}

All ASR experiments were performed using the Kaldi ASR toolkit \cite{povey2011kaldi} and the data described in Section~\secref{SEC:corpus}. 
For  multilingual  training, the training sets of all the relevant languages were pooled.
No phone merging was performed between languages and hence all acoustic models are language dependent.

Context-dependent Gaussian mixture model - hidden Markov models (GMM-HMM) were trained to obtain the alignments required for neural network training. 
Three-fold data augmentation~\cite{ko2015audio} was applied prior to feature extraction for neural network training.
The feature set used to train the neural network comprised MFCCs (40-dimensional, without derivatives), pitch features (3-dimensional) and i-vectors for speaker adaptation (100-dimensional)~\cite{saon2013speaker}.

The acoustic models of all ASR systems were trained according the standard Kaldi CNN-TDNN-F~\cite{povey2018} Librispeech recipe (6 CNN layers and 10 time-delay layers followed by a rank reduction layer) using the default hyperparameters.
For the bilingual experiments, the multilingual acoustic models were subsequently adapted to the four different target language pairs. 

\begin{figure} [t]
	\centering
	\includegraphics[width=\columnwidth]{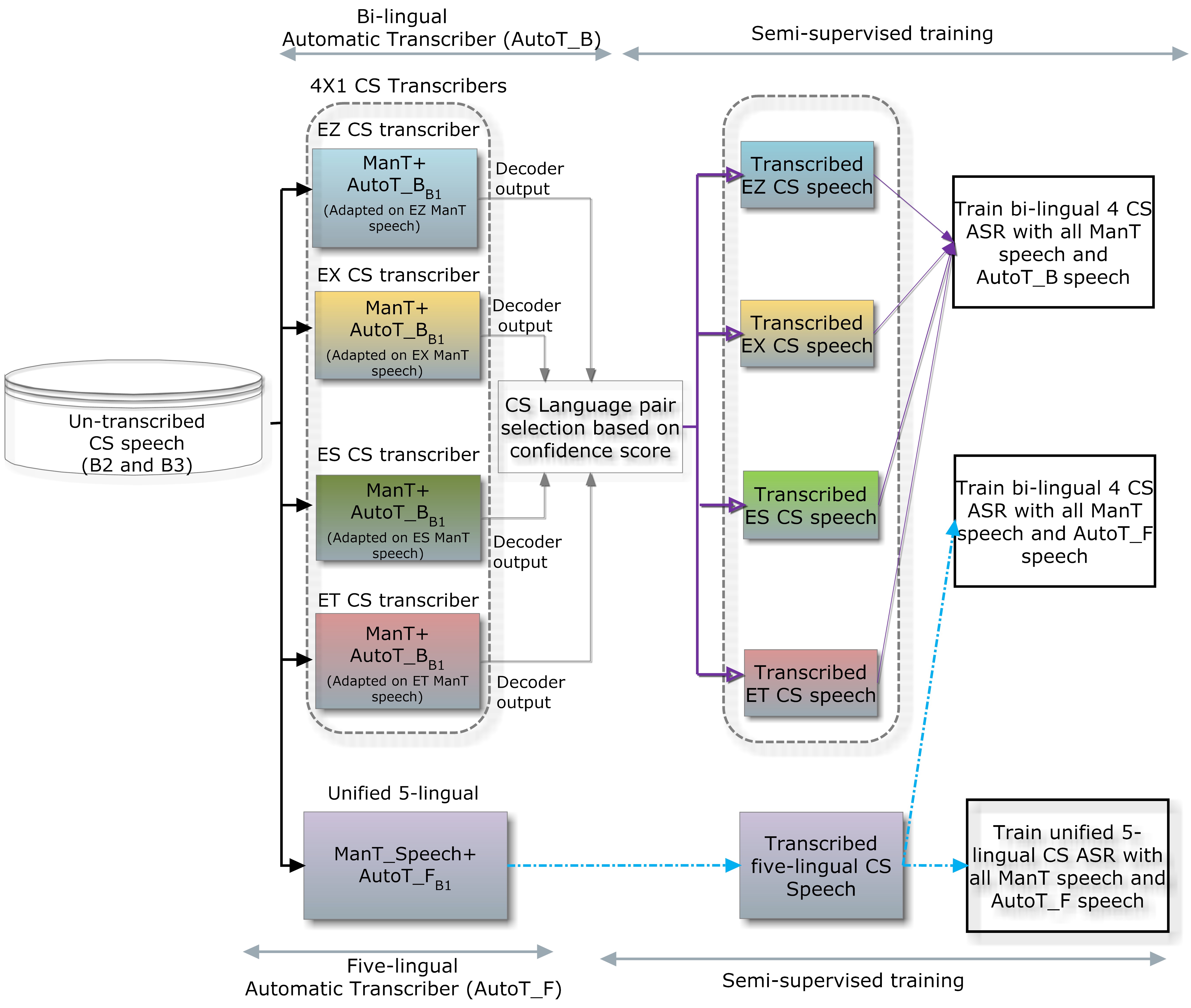}
	\caption{Semi-supervised training framework for bilingual and five-lingual systems.}
	\label{mdnn}
\end{figure}

\section{Automatic transcription systems}
A recent study demonstrated that semi-supervised training can improve the performance of Frisian-Dutch code-switched ASR~\cite{yilmaz2018semi}.
A similar approach was taken in this study, using the system configuration shown in Figure~\ref{mdnn}. 
The figure illustrates the two phases of semi-supervised training for the parallel bilingual as well as five-lingual configurations: automatic transcription followed by acoustic model retraining. 

\subsection{Parallel bilingual transcription}
The first transcription system in Figure~\ref{mdnn} consists of four subsystems, each corresponding to a code-switch language pair~(4$\times$CS in Figure~\ref{mdnn}). 
Acoustic models were trained on the manually transcribed soap opera data (ManT) described in Section~\secref{SEC:corpus:transcribed}\ pooled with the automatically transcribed speech (AutoT\_B) introduced in Section~\secref{SEC:corpus:untranscribed1}.
Because the languages in the untranscribed data were unknown, each utterance was decoded in parallel by each of the bilingual decoders.
The output with the highest confidence score provided both the transcription and a language pair label for each segment.

\subsection{Five-lingual transcription}
The second transcription system was based on a single acoustic model trained on all five languages.
The training data consisted of the manually transcribed soap opera speech (ManT) pooled with the transcriptions generated by a five-lingual system (AutoT\_F).
Since the five-lingual system is not restricted to bilingual output, Bantu-to-Bantu language switching was possible and was observed in these transcriptions.
Moreover, the automatically generated transcriptions sometimes contained more than two languages. Although the use of more than two languages within a single utterance is not common, our soap opera data does include a few such examples.
%
%
\section{Language Modelling}
\label{sec:typestyle}
\subsection{Baseline language model}
The EZ, EX, ES, ET vocabularies contained 11\,292, 8\,805, 4\,233, 4\,957 word types respectively and were closed with respect to the training, development and test sets.
The SRILM toolkit \cite{stolcke2002srilm} was used to train and evaluate all language models (LMs).
Four bilingual and one five-lingual trigram language model were used for the transcription systems as well as for semi-supervised training~\cite{yilmaz2018semi,biswas2019IS2}.
Table~\ref{perplexity} summarises the development and test set perplexities.
Details on the monolingual and code-switch perplexities are only provided for the test set (columns 3 to 6 in Table ~\ref{perplexity}).

Much more monolingual English text was available for language model development than text in the Bantu languages (471M vs 8M words).
Therefore, the monolingual perplexity (MPP) is much higher for the Bantu languages than for English for each language pair.

Code-switch perplexities (CPP) for language switches indicate the uncertainty of the first word following a language switch. 
EB corresponds to switches from English to a Bantu language and BE indicates a switch in the other direction. 
Table~\ref{perplexity} shows that the CPP for switching from English to isiZulu and isiXhosa is much higher than switching from these languages to English. 
This can be ascribed to the much larger isiZulu and isiXhosa vocabularies, which are, in turn, due to the high degree of agglutination and the use of conjunctive orthography in these languages.
The CPP values are even higher for the five-lingual language model.
This is because the five-lingual trigrams allow language switches not permitted by the bilingual models.

\begin{table}[t]
\scriptsize
\centering
\begin{tabular*}{0.47\textwidth}{@{\extracolsep{\fill}} l @{\hspace{4pt}} r @{\hspace{4pt}} r @{\hspace{4pt}} r @{\hspace{4pt}} r @{\hspace{4pt}} r @{\hspace{4pt}} r @{\hspace{4pt}} r @{\hspace{4pt}} r @{} }
\toprule
 & Dev & Test & all CPP & CPP$_{\rm EB}$ & CPP$_{\rm BE}$ & all MPP & MPP$_{\rm E}$ & MPP$_{\rm Z}$ \\
\midrule
\multicolumn{9}{c}{\textbf{Bilingual trigram language model}} \\
EZ & 425.8 & 601.7 & 3\,291.9 & 3\,835.0 & 2\,865.4 & 358.1 & 121.1 & 777.8 \\
EX & 352.9 & 788.8 & 4\,914.4 & 6\,549.6 & 3\,785.6 & 459.0 & 96.8 & 1\,355.6 \\ 
ES & 151.5 & 180.5 & 959.0 & 208.6 & 4\,059.1 & 121.2 & 126.9 & 117.8 \\ 
ET & 213.3 & 224.5 & 70.2 & 317.3 & 3\,798.1 & 160.4 & 142.1 & 176.1 \\
\midrule
\multicolumn{9}{c}{\textbf{Five-lingual trigram language model}} \\
EZ & 599.9 & 1\,007.1 & 6\,708.2 & 17\,371.0 & 2\,825.2 & 561.8 & 94.4 & 2\,013.0 \\
EX & 669.1 & 1\,881.9 & 15\,083.6 & 50\,208.3 & 5\,058.0 & 1\,015.9 & 87.6 & 5\,590.0 \\
ES & 365.5 & 345.3 & 3\,617.4 & 2\,607.1 & 5\,088.8 & 207.8 & 103.9 & 355.8 \\
ET & 237.0 & 277.5 & 2\,936.6 & 1\,528.4 & 5\,446.3 & 158.1 & 99.8 & 211.2 \\
\bottomrule
\end{tabular*}%
\caption{Development and test set perplexities. CPP: code-switch perplexity. MPP: monolingual perplexity.}
\label{perplexity}
\end{table}

\subsection{Semi-supervised language models}
Only bilingual transcriptions were considered for the semi-supervised language model experiments.
The automatically generated transcriptions of data sets B2 and B3 were added to the language model training data, similar to the approach proposed in~\cite{drugman2019active}.
Semi-supervised bilingual language models for each language pair were obtained by interpolating the baseline trigram with a trigram derived from the text in the transcriptions of the corresponding target language pair. 
The interpolation weights were optimised using the development data. 

A related study has shown that text data augmentation can be useful in under-resourced scenarios~\cite{emre2018IS}. 
This approach was evaluated on the EZ subset of our data by training a long short-term memory (LSTM) model on the manual and automatic transcriptions to generate additional artificial text data.
The model was subsequently used to generate a data set of approximately 11.5 million words.
The semi-supervised English-isiZulu language model described in the previous paragraph was interpolated with a trigram trained on this artificial data set. 

In a further attempt to strengthen the language model at code-switch points,  1 million of artificial code-switched trigrams were synthesised using the method described in \cite{van2018synthesised}.
The perplexity values of the resulting language models are reported in Table \ref{tab:perplexity2}.

The first row in the table shows that adding the transcriptions of the automatically transcribed data to the LM training set reduces the test set perplexity of the EZ semi-supervised language model by more than 50 relative to the baseline value in Table~\ref{perplexity}.  
The ET semi-supervised language model also achieved a significant perplexity reduction on the test set. 
However, the EX and ES semi-supervised language models did not show any improvement compared to their respective baselines. 
This may be because there are far fewer isiXhosa and Sesotho segments in the automatically generated transcriptions than isiZulu and Setswana segments (cf. Table~\ref{tab:Trans_details}).

\vspace*{2mm}
\begin{table}[h]
\huge
\renewcommand*{\arraystretch}{1.2}
\resizebox{\columnwidth}{!}{%
\begin{tabular}{llrrrrrrrr}
 \toprule[1pt]\midrule[0.3pt]
 & Resources & \multicolumn{1}{l}{Dev} & \multicolumn{1}{l}{Test} & \multicolumn{1}{l}{all CPP} & \multicolumn{1}{l}{CPP$_{EB}$} & \multicolumn{1}{l}{CPP$_{BE}$} & \multicolumn{1}{l}{all MPP} & \multicolumn{1}{l}{MPP$_{E}$} & \multicolumn{1}{l}{MPP$_{Z}$} \\ \toprule[1pt]\midrule[0.3pt]
\multirow{3}{*}{EZ} & \begin{tabular}[c]{@{}l@{}}ManT + AutoT\_B$_{\rm B1}$\\ + AutoT\_B$_{\rm B2B3}$\end{tabular} & 392.2 & 547.4 & 2898.8 & 3297.6 & 2578.3 & 328.5 & 108.2 & 727.4 \\ \cline{2-10} 
 & + 11.5M artificial text & 362.8 & 507.5 & 2368.8 & 3005.8 & 1907.7 & 315.9 & 103.7 & 701.3 \\ \cline{2-10} 
 & + 1M synthetic bigrams & 358.3 & 501.9 & 2139.8 & 2613.5 & 1784.2 & 320.7 & 108.3 & 697.5 \\ \midrule
EX & \begin{tabular}[c]{@{}l@{}}ManT + AutoT\_B$_{\rm B1}$\\ + AutoT\_B$_{\rm B2B3}$\end{tabular} & 345.4 & 787.4 & 5039.1 & 7176.3 & 3654.6 & 454.5 & 89.2 & 1411.4 \\ \midrule
ES & \begin{tabular}[c]{@{}l@{}}ManT + AutoT\_B$_{\rm B1}$\\ + AutoT\_B$_{\rm B2B3}$\end{tabular} & 200.7 & 206.7 & 1074.3 & 347.0 & 3487.7 & 144.9 & 117.3 & 170.8 \\ \midrule
ET & \begin{tabular}[c]{@{}l@{}}ManT + AutoT\_B$_{\rm B1}$\\ + AutoT\_B$_{\rm B2B3}$\end{tabular} & 138.4 & 164.8 & 938.0 & 214.6 & 3784.7 & 108.8 & 105.5 & 111.0 \\ \bottomrule[1pt]\midrule[0.3pt]
\end{tabular}
}
\caption{Development and test set perplexities for different language model configurations. AutoT\_B$_{\rm B1}$: B1 transcribed as described in~\protect\cite{biswas2019IS2}. AutoT\_B$_{\rm B2B3}$: B2 transcribed by system \textbf{A} and B3 transcribed by system \textbf{C} (Figure~\ref{SST_stra}).}
\label{tab:perplexity2}
\end{table}


Table~\ref{tab:perplexity2} also shows that the additional text generated by the LSTM reduced the EZ development and test perplexity values substantially. 
The additional text also helped to bolster the language model at code-switch points. 
EZ CPP improved further after the 1M synthesized trigrams were added.
%
%
\section{Experiments}
\begin{figure} [t]
	\centering
	\includegraphics[scale=0.04]{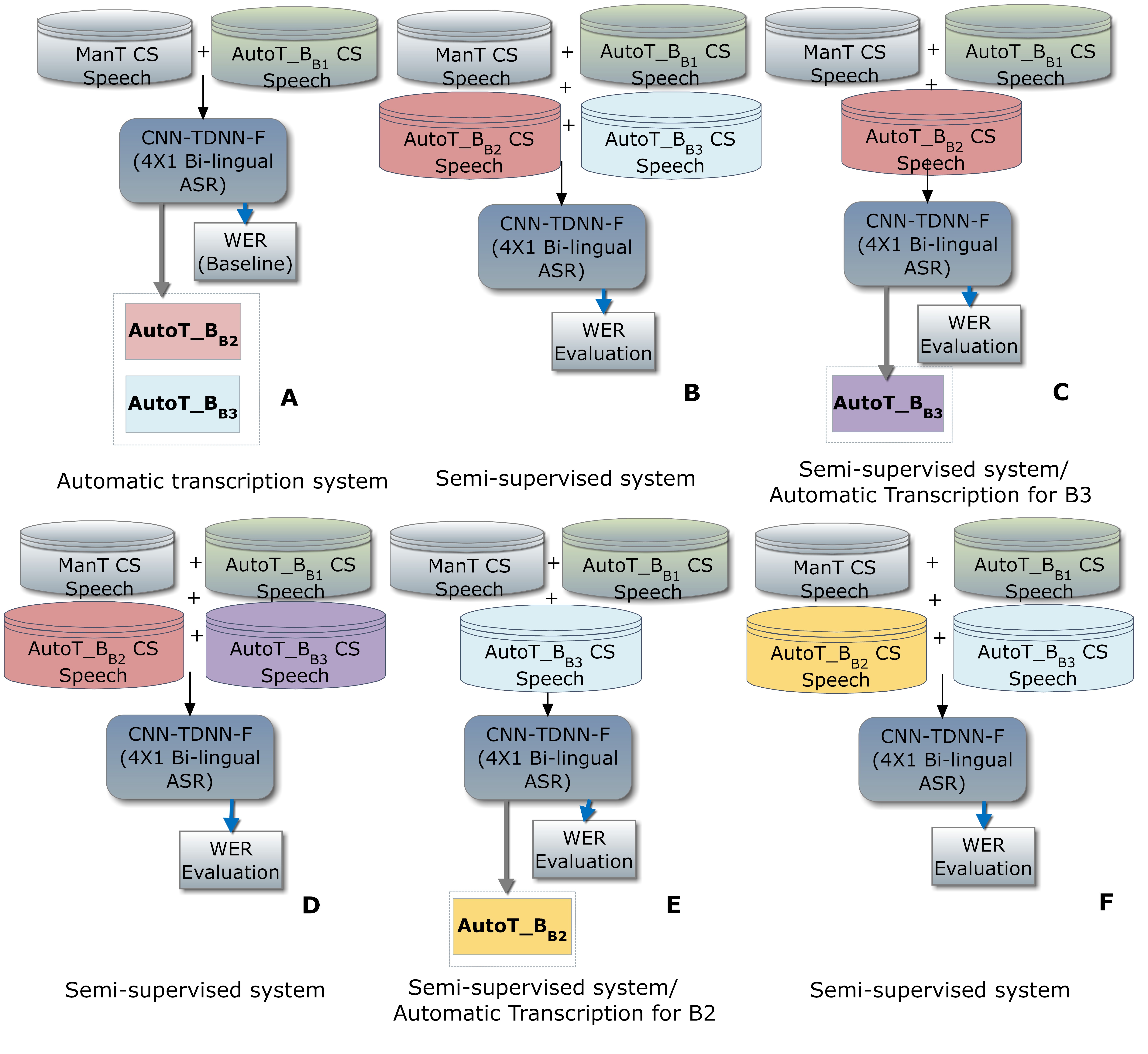}
	\caption{Semi-supervised training configurations for bilingual ASR. System names are given in parenthesis. (A:~Baseline, B: Without batches, C to F: Batchwise)}
	\label{SST_stra}
\end{figure}

Bilingual semi-supervised acoustic model training experiments were performed using batches B2 and B3 according to the two approaches illustrated in Figure \ref{SST_stra}.
Similar configurations were used for the five-lingual experiments. In the first approach,
both batches were first automatically transcribed using the baseline ASR (System \textbf{A} in Figure~\ref{SST_stra}) followed by retraining using automatic transcriptions for both batches (System \textbf{B}).

The second approach used batch-wise semi-supervised training. 
Using System \textbf{A}, B2 was transcribed first, followed by acoustic model retraining  with the  automatically transcribed B2 data included in the training set (System \textbf{C}).
B3 was then transcribed using the updated models and the retraining process repeated, this time also including the transcriptions of B3 (System \textbf{D}).
This order was also reversed i.e. B3 first (System \textbf{E}) followed by B2 (System \textbf{F}). 

The experimental procedure for the bilingual and five-lingual systems is summarised in Table~\ref{tab:SST_systems}.
The manual transcriptions introduced in Section~\secref{SEC:corpus:transcribed}\ were always included in the training set.
The composition of the automatic transcriptions included during training is shown in the last three columns of the table, with the last two indicating which systems were used to generate transcriptions for B2 and B3 respectively. 
Preliminary experiments indicated that the bilingual ASR systems achieved best performance when trained on AutoT\_F transcriptions \cite{biswas2019IS2}. 
Thus, the bilingual systems considered here were also trained on the AutoT\_F transcriptions of B2 and B3 (System N in Table~\ref{tab:SST_systems}).

\begin{table}[h!]
\centering
\footnotesize
\renewcommand*{\arraystretch}{0.8}
\resizebox{.9\columnwidth}{!}{%
\begin{tabular}{@{}llllll@{}}
\toprule
\multicolumn{1}{c}{\multirow{2}{*}{Type of ASR}} & \multicolumn{1}{c}{\multirow{2}{*}{System}} & \multicolumn{1}{c}{\multirow{2}{*}{}} & \multicolumn{3}{c}{AutoT} \\ \cmidrule(l){4-6} 
\multicolumn{1}{c}{} & \multicolumn{1}{c}{} & \multicolumn{1}{c}{} & \multicolumn{1}{c}{B1} & \multicolumn{1}{c}{B2} & \multicolumn{1}{c}{B3} \\ \midrule
\multirow{6}{*}{\begin{tabular}[c]{@{}l@{}}Bilingual\end{tabular}} & A (Baseline) & \multirow{6}{*}{} & \multirow{6}{*}{AutoT\_B} & - & - \\
 & B &  &  & \begin{tabular}[c]{@{}l@{}}A\end{tabular} & \begin{tabular}[c]{@{}l@{}}A\end{tabular} \\
 & C &  &  & \begin{tabular}[c]{@{}l@{}}A\end{tabular} & - \\
 & D &  &  & \begin{tabular}[c]{@{}l@{}}A\end{tabular} & \begin{tabular}[c]{@{}l@{}}C\end{tabular} \\
 & E &  &  & - & \begin{tabular}[c]{@{}l@{}}A\end{tabular} \\
 & F &  &  & \begin{tabular}[c]{@{}l@{}}E\end{tabular} & \begin{tabular}[c]{@{}l@{}}A\end{tabular} \\ \midrule
\multirow{6}{*}{\begin{tabular}[c]{@{}l@{}}5-lingual\end{tabular}} & G (Baseline) & \multirow{6}{*}{} & \multirow{6}{*}{AutoT\_F} & - & - \\
 & H &  &  & \begin{tabular}[c]{@{}l@{}}G\end{tabular} & \begin{tabular}[c]{@{}l@{}}G\end{tabular} \\
 & I &  &  & \begin{tabular}[c]{@{}l@{}}G\end{tabular} & - \\
 & J &  &  & \begin{tabular}[c]{@{}l@{}}G\end{tabular} & \begin{tabular}[c]{@{}l@{}}I\end{tabular} \\
 & K &  &  & - & \begin{tabular}[c]{@{}l@{}}G\end{tabular} \\
 & L &  &  & \begin{tabular}[c]{@{}l@{}}K\end{tabular} & \begin{tabular}[c]{@{}l@{}}G\end{tabular} \\ \midrule
\multicolumn{1}{c}{\multirow{2}{*}{\begin{tabular}[c]{@{}l@{}}Bilingual\\ (5-lingual trans.)\end{tabular}}} & \multirow{2}{*}{N} & \multirow{2}{*}{} & \multirow{2}{*}{AutoT\_F} & \multirow{2}{*}{\begin{tabular}[c]{@{}l@{}}G\end{tabular}} & \multirow{2}{*}{\begin{tabular}[c]{@{}l@{}}I\end{tabular}} \\
\multicolumn{1}{c}{} &  &  &  &  &  \\\bottomrule
\end{tabular}%
}
\caption{Semi-supervised acoustic model configurations.}
\label{tab:SST_systems}
\end{table}
%
%
\section{Results and Discussion}
\label{ssec:subhead}

\subsection{Automatic transcription}
The output of the transcription systems is summarised in Table \ref{tab:Trans_details}. 
The first five rows in the table correspond to segments that were classified as monolingual while the last row shows the number of segments that contain code-switching.
The values in this row reveal a high number of code-switched segments in data sets B2 and B3. 
It should be kept in mind that transcriptions of mixed language utterances produced by configurations A, C and E can only contain code-switching between English and one Bantu language.
In contrast, the utterances produced by G, I and K can contain examples of code-switching between two or more Bantu languages.

\begin{table}[h!]
\large
\renewcommand*{\arraystretch}{1.0}
\resizebox{\columnwidth}{!}{%
\begin{tabular}{@{}lrrrrrrrr@{}}
\toprule
\multirow{2}{*}{Language} & \multicolumn{2}{c}{A} & \multicolumn{1}{c}{C} & \multicolumn{1}{c}{E} & \multicolumn{2}{c}{G} & \multicolumn{1}{c}{I} & \multicolumn{1}{c}{K} \\ \cmidrule(lr){2-3} \cmidrule(lr){4-4} \cmidrule(lr){5-5} \cmidrule(lr){6-7} \cmidrule(lr){8-8} \cmidrule(l){9-9} 
 & \multicolumn{1}{c}{B2} & \multicolumn{1}{c}{B3} & \multicolumn{1}{c}{B3} & \multicolumn{1}{c}{B2} & \multicolumn{1}{c}{B2} & \multicolumn{1}{c}{B3} & \multicolumn{1}{c}{B3} & \multicolumn{1}{c}{B2} \\ \midrule
English & 8\,570 & 11\,746 & 12\,027 & 8\,794 & 12\,373 & 16\,333 & 17\,059 & 12\,898 \\
isiZulu & 5\,955 & 6\,190 & 6\,854 & 6\,604 & 4\,708 & 4\,587 & 4\,641 & 4\,496 \\
isiXhosa & 302 & 244 & 324 & 337 & 332 & 72 & 108 & 174 \\
Sesotho & 1\,317 & 1\,310 & 1043 & 1\,889 & 1\,067 & 678 & 581 & 797 \\
Setswana & 2\,598 & 2\,602 & 3\,042 & 2\,539 & 1\,244 & 1\,009 & 1~038 & 1\,124 \\
Code-switched & 25\,824 & 28\,573 & 27\,407 & 24\,257 & 24\,842 & 27\,675 & 27\,077 & 25\,086  \\ \bottomrule
\end{tabular}%
}
\caption{Number of segments per language for different transcription systems.}
\label{tab:Trans_details}
\end{table}

\subsection{Automatic speech recognition}
ASR quality was measured in terms of word error rate (WER) evaluated on the development and test sets for each language pair described in Table~\ref{tab:corpora_stat}.
Results for the different semi-supervised training configurations are reported in Table~\ref{tab:result1}.
 
 \begin{table}[h!]
 \large
\resizebox{\columnwidth}{!}{%
\begin{tabular}{@{}lrrrrrrrrrrrr@{}}
\toprule
\multicolumn{13}{c}{\textbf{Bilingual ASR}} \\
\multicolumn{1}{c}{} & \multicolumn{2}{c}{A (Baseline)} & \multicolumn{2}{c}{B} & \multicolumn{2}{c}{C} & \multicolumn{2}{c}{D} & \multicolumn{2}{c}{E} & \multicolumn{2}{c}{F} \\ \cmidrule(lr){2-3} \cmidrule(lr){4-5} \cmidrule(lr){6-7}\cmidrule(lr){8-9}\cmidrule(lr){10-11} \cmidrule(lr){12-13}
\multicolumn{1}{c}{\multirow{-2}{*}{\begin{tabular}[c]{@{}c@{}}CS\\ Pair\end{tabular}}} & \multicolumn{1}{c}{Dev} & \multicolumn{1}{c}{Test} & \multicolumn{1}{c}{Dev} & \multicolumn{1}{c}{Test} & \multicolumn{1}{c}{Dev} & \multicolumn{1}{c}{Test} & \multicolumn{1}{c}{Dev} & \multicolumn{1}{c}{Test} & \multicolumn{1}{c}{Dev} & \multicolumn{1}{c}{Test} & \multicolumn{1}{c}{Dev} & \multicolumn{1}{c}{Test} \\ \midrule
EZ & 34.5 & 40.8 & 33.4 & 39.3 & 32.7 & 39.6 & 32.8 & 38.6 & 34.6 & 39.9 & 32.8 & 39.0 \\
EX & 35.8 & 42.7 & 34.8 & 41.8 & 35.5 & 42.0 & 35.0 & 41.0 & 33.7 & 41.4 & 34.3 & 42.2 \\
ES & 51.6 & 48.7 & 49.3 & 47.3 & 49.0 & 46.5 & 48.7 & 45.8 & 49.5 & 47.4 & 49.5 & 46.4 \\
ET & 44.3 & 41.3 & 41.8 & 39.9 & 41.8 & 40.3 & 40.7 & 39.6 & 42.3 & 39.3 & 42.2 & 38.6 \\
\rowcolor[HTML]{FFFFC7} 
Overall & 41.5 & 43.4 & 39.8 & 42.1 & 39.8 & 42.1 & 39.3 & \textbf{41.3} & 40.0 & 42.0 & 39.7 & 41.6 \\ \midrule
\multicolumn{13}{c}{\textbf{5-lingual ASR}} \\
\multicolumn{1}{c}{} & \multicolumn{2}{c}{G (Baseline)} & \multicolumn{2}{c}{H} & \multicolumn{2}{c}{I} & \multicolumn{2}{c}{J} & \multicolumn{2}{c}{K} & \multicolumn{2}{c}{L} \\ \cmidrule(lr){2-3} \cmidrule(lr){4-5} \cmidrule(lr){6-7}\cmidrule(lr){8-9}\cmidrule(lr){10-11} \cmidrule(lr){12-13}
\multicolumn{1}{c}{\multirow{-2}{*}{\begin{tabular}[c]{@{}c@{}}CS\\ Pair\end{tabular}}} & \multicolumn{1}{c}{Dev} & \multicolumn{1}{c}{Test} & \multicolumn{1}{c}{Dev} & \multicolumn{1}{c}{Test} & \multicolumn{1}{c}{Dev} & \multicolumn{1}{c}{Test} & \multicolumn{1}{c}{Dev} & \multicolumn{1}{c}{Test} & \multicolumn{1}{c}{Dev} & \multicolumn{1}{c}{Test} & \multicolumn{1}{c}{Dev} & \multicolumn{1}{c}{Test} \\ \midrule
EZ & 37.6 & 43.6 & 36.3 & 41.0 & 35.7 & 42.5 & 34.3 & 42.1 & 35.7 & 42.5 & 35.0 & 42.0 \\
EX & 40.6 & 54.5 & 37.4 & 50.0 & 37.7 & 50.7 & 38.3 & 49.0 & 37.7 & 50.7 & 37.9 & 48.5 \\
ES & 54.5 & 49.3 & 51.5 & 48.1 & 52.8 & 46.9 & 51.5 & 47.8 & 52.8 & 49.9 & 51.6 & 48.0 \\
ET & 47.2 & 43.9 & 46.1 & 42.3 & 46.4 & 42.4 & 44.1 & 40.9 & 46.4 & 42.4 & 45.6 & 41.6 \\
\rowcolor[HTML]{FFFFC7} 
Overall & 46.5 & 46.7 & 44.3 & 44.4 & 44.8 & 44.8 & 43.5 & \textbf{44.2} & 44.8 & 44.8 & 44.1  & 44.3 \\ \bottomrule
\end{tabular}%
}
 \caption{Mixed WERs (\%) for the four code-switched language pairs evaluated using the baseline language model.}
 \label{tab:result1}
\end{table}
 
\subsubsection{Bilingual semi-supervised training}
\label{sssec:subsubhead}
The upper part of Table~\ref{tab:result1} shows that, on average, semi-supervised training using 53 additional hours of speech data (B) yields an absolute improvement of 1.3\% over the baseline (A) on the test set. 
It could be argued that this improvement is not large, given how much additional data was added to the training set. 
However, the segments were not created by language experts and may therefore not be accurate. 
Improving the quality of the segments might lead to better performance.
The overall WER values for systems D and F show that batch-wise training results in better performance than processing all the untranscribed data in a single step.
The best performing bilingual semi-supervised system (D) achieved an absolute overall WER improvement of 2.1\% over the baseline on the test set.

\subsubsection{Five-lingual semi-supervised training}
The lower half of Table~\ref{tab:result1} indicates that the five-lingual semi-supervised acoustic models also benefited from the additional data.
As for the bilingual systems, the five-lingual system yielded better results when using batch-wise training.
The best performing system (J) outperformed the baseline by 2.5\% absolute on the test set.
Although the WER achieved by the five-lingual system is higher than that achieved by the bilingual system, this remains a promising result.
Five-lingual recognition is more difficult since it allows more freedom in terms of the permissible language switches.
It does, however, more honestly reflect the large and undetermined mix of languages an ASR system may be confronted with when processing South African speech.

\subsubsection{Bilingual semi-supervised training with five-lingual transcriptions \& semi-supervised language model}
The bilingual acoustic models retrained with transcriptions generated by the five-lingual system (N) achieved the best overall WER: 40.56\% which is an absolute improvement of 0.7\% over system D, its closest competitor. 
This improvement may be due to the five-lingual system's ability to transcribe in more than two languages, since the untranscribed soap opera speech is known to contain at least some utterances that do not conform to the four bilingual systems. 
System N was also evaluated in combination with the semi-supervised language model.
The combination of the semi-supervised acoustic and language models (N$_{LM1}$) reduced the overall WER on the test set by another 0.5\% absolute.  

Due to computational constraints, additional language model experiments were only conducted on the EZ data set.
Table~\ref{tab:result_lm_arti} shows that the use of language models derived from the additional text generated by the LSTM model, N$_{LM2}$, and the synthesised trigrams including additional text generated by the LSTM model, N$_{LM3}$, resulted in further small reductions in WER for both the development and test sets. 

Moreover, the corresponding code-switch bigram accuracy (row 2 in Table~\ref{tab:result_lm_arti}) also improved when the language model training data included the additional, artificially generated text and trigrams.

 \begin{table}[!h]
 \scriptsize
\centering
\renewcommand*{\arraystretch}{0.7}
\resizebox{.9\columnwidth}{!}{%
\begin{tabular}{@{}ccccccc@{}}
\toprule
\multirow{2}{*}{} & \multicolumn{2}{c}{N$_{LM1}$} & \multicolumn{2}{c}{N$_{LM2}$} & \multicolumn{2}{c}{N$_{LM3}$} \\ \cmidrule(lr){2-3} \cmidrule(lr){4-5} \cmidrule(lr){6-7} 
 & Dev & Test & Dev & Test & Dev & Test \\ \midrule
\multicolumn{1}{l}{WER} & \multicolumn{1}{r}{31.3} & \multicolumn{1}{r}{36.9} & \multicolumn{1}{r}{30.3} & \multicolumn{1}{r}{36.7} & \multicolumn{1}{r}{30.5} & \multicolumn{1}{r}{\textbf{36.2}} \\ 
\multicolumn{1}{l}{CSBA} & \multicolumn{1}{r}{42.3} & \multicolumn{1}{r}{40.1} & \multicolumn{1}{r}{45.1} & \multicolumn{1}{r}{41.0} & \multicolumn{1}{r}{45.9} & \multicolumn{1}{r}{\textbf{42.1}} \\\bottomrule
\end{tabular}%
}
 \caption{Mixed WER (\%) and code-switched bigram accuracy (CSBA) (\%) for EZ using artificial and synthetic text.}
 \label{tab:result_lm_arti}
\end{table}
\subsection{Language specific WER analysis}
For code-switched ASR, the performance of the recogniser at the code-switch points is of particular interest. 
Language specific WERs and code-switched bigram accuracy values for the different systems are presented in Table \ref{tab:detail_analysis}.
All values are percentages.

\begin{table}[h!]
\huge
\resizebox{\columnwidth}{!}{%
\begin{tabular}{@{}lllllllllllll@{}}
\toprule
\multirow{2}{*}{System} & \multicolumn{3}{c}{EZ} & \multicolumn{3}{c}{EX} & \multicolumn{3}{c}{ES} & \multicolumn{3}{c}{ET} \\ \cmidrule(lr){2-4} \cmidrule(lr){5-7} \cmidrule(lr){8-10} \cmidrule(lr){11-13}
 & \multicolumn{1}{c}{E} & \multicolumn{1}{c}{Z} & \multicolumn{1}{c}{CSBA} & \multicolumn{1}{c}{E} & \multicolumn{1}{c}{X} & \multicolumn{1}{c}{CSBA} & \multicolumn{1}{c}{E} & \multicolumn{1}{c}{S} & \multicolumn{1}{c}{CSBA} & \multicolumn{1}{c}{E} & \multicolumn{1}{c}{T} & \multicolumn{1}{c}{CSBA} \\ \midrule
A (baseline) & 37.9 & 48.7 & 33.3 & 37.8 & 54.5 & 25.8 & 43.7 & 61.4 & 25.2 & 36.2 & 51.8 & 35.6 \\
D & 32.1 & 43.6 & 39.8 & 31.3 & 48.3 & 33.9 & 33.0 & 55.9 & 34.8 & 27.6 & 47.6 & 42.3 \\
G (baseline) & 29.6  & 54.3 & 16.3 & 34.1  & 70.1  & 7.3 & 29.3  & 65.2  & 7.5 & 23.9 & 57.16 & 11.3 \\
J & 28.2 & 52.7 & 16.6  & 29.2 & 64.0 & 11.7 & 28.8  & 62.9 & 7.7 & 21.5 & 53.8 & 13.8 \\
N & 30.3 & 42.9 & 40.2 & 30.3 & 47.2 & 32.8 & 33.6 & 55.8 & 35.5 & 28.1 & 46.0 & 42.2 \\
N$_{LM1}$ & 29.0 & 42.9 & 40.1 & 29.9 & 47.0 & 32.7 & 32.2 & 55.8 & 35.8 & 26.3 & 46.1 & 43.6 \\
\bottomrule
\end{tabular}%
}
\caption{Language specific WER (\%)  for English (E), isiZulu (Z), isiXhosa (X), Sesotho (S), Setswana (T) and code-switched bigram accuracy (CSBA) (\%) for the test set.}
\label{tab:detail_analysis}
\end{table}

The table reveals that five-lingual ASR (systems G \& J) is more biased towards English (lower WER) than the bilingual systems (systems A \& D), for which the Bantu language WERs are better. 
As already observed for the language model, the bias of the five-lingual system towards English is due to the much larger proportion of in-domain English training material available when pooling the four code-switched language pairs.

The accuracy at the code-switch points is substantially better when using the bilingual semi-supervised system.
This is probably due to the higher ambiguity encountered by the five-lingual system at code-switch points than the bilingual systems.
The table also reveals that the improvements observed for systems using the semi-supervised language models are due mostly to improved English recognition (N$_{LM1}$). 

%
%
\section{Conclusions}
In this study we evaluated semi-supervised acoustic and language model training with the aim of improving the ASR performance of under-resourced code-switched South African speech. 
Two batches (approximately 53 hours in total) of manually segmented but untranscribed soap opera speech, rich in code-switching, were used for experimentation.
The new speech was processed both in a single step and in batches by two automatic transcription systems: one comprising four parallel bilingual recognisers and the other a single five-lingual system. 

The results indicate that the overall WER of both bilingual and five-lingual systems was reduced substantially and that batch-wise training was the better approach in both instances.
However, the overall average performance of the bilingual systems remained better than that of the five-lingual system. 
This is probably because the five-lingual recognition task is inherently more complex. 

The five-lingual system exhibited a bias towards English while the four bilingual recognisers were more accurate for the Bantu languages.
Despite the confusability inherent in decoding five languages, the five-lingual system achieved an error rate that was almost as good as that attained by the bilingual systems. 
Thus, it seems worthwhile developing bilingual and five-lingual code-switched ASR systems in parallel. 

The semi-supervised language model resulted in a significant reduction in perplexity that translated into a corresponding decrease in WER. 
Artificially generated code-switched text and synthetic trigrams also showed potential to further improve the ASR performance.  

Future work will focus on incorporating automatic segmentation as well as speaker and language diarisation to extend the pool of training data.
The effect of the number and size of training batches on system performance will also be studied more carefully. 
%
%
\section{Acknowledgements}
We would like to thank the Department of Arts \& Culture (DAC) of the South African government for funding this research. 
We are grateful to e.tv and Yula Quinn at Rhythm City, as well as the SABC and Human Stark at Generations: The Legacy, for assistance with data compilation. 
We also gratefully acknowledge the support of the NVIDIA corporation for the donation of GPU equipment.

\section{References}

\bibliographystyle{lrec}
\bibliography{astik_refer_new}



\end{document}